\documentclass[%
 reprint,
 amsmath,amssymb,
 aps,
]{revtex4-2}

\usepackage{graphicx}
\usepackage{dcolumn}
\usepackage{bm}
\usepackage{rotating}
\usepackage{filecontents}
\begin{document}

\preprint{APS/123-QED}

\title{Electric monopole transitions and low-lying level structure in $^{106}$Pd}

\author{N.~Marchini$^{1,2}$, A.~Nannini$^2$, M.~Ottanelli$^2$, A.~Saltarelli$^{1,3}$, G.~Benzoni$^4$, E.~R.~Gamba$^{5*}$, A.~Goasduff$^6$, A. Gottardo$^6$, J.~Ha$^6$, T.~Krings$^7$, M.~Perri$^{1,3}$, M.~Polettini$^{4,8}$, M.~Rocchini$^9$, P.~Sona$^2$}
\affiliation{$^1$Universit\`a degli Studi di Camerino, Dipartimento di Fisica, IT-62032 Camerino, Italy \\
$^2$INFN Sezione di Firenze, IT-50019 Firenze, Italy\\
$^3$INFN Sezione di Perugia, IT-06123 Perugia, Italy\\
$^4$INFN Sezione di Milano, IT-20133 Milano, Italy\\
$^5$Museo Storico della Fisica e Centro Studi e Ricerche Enrico Fermi, IT-00184 Roma, Italy\\
$^6$INFN Laboratori Nazionali di Legnaro, IT-35020 Padova, Italy\\
$^7$Institut f\"{u}r Kernphysik (IKP), 52428 J\"{u}lich, Germany\\
$^8$Universit\`a degli Studi di Milano, Dipartimento di Fisica, IT-20133 Milano, Italy \\
$^9$University of Guelph, Department of Physics, N1G2W1 Guelph, Canada\\
$^*$ Current affiliation: Dipartimento di Fisica, Università degli Studi di Milano and INFN, 20133 Milano, Italy.}
\date{\today}

\begin{abstract}
The structure of $^{106}$Pd, populated in the EC/$\beta^+$ decay of $^{106}$Ag, was investigated at the INFN Legnaro National Laboratories using the Spes Low-energy Internal Conversion Electron Spectrometer (SLICES). The K-internal conversion coefficients of some transitions have been measured and the electric monopole transition strengths, between low-lying 2$^+$ and 0$^+$ states, have been deduced. These experimental data combined with the results from internal conversion electron measurement on $^{104}$Pd previously performed by the Florence spectroscopy group were compared with the theoretical values calculated in the framework of the interacting proton-neutron boson model. Good agreement between experimental and theoretical values is found when interpreting the 0$^+_3$ level as an intruder state.
\end{abstract}

\maketitle

\section{Introduction}
The collective properties of stable Pd isotopes (Z=46) have been
the focus of several experimental and theoretical studies in
the past decades. 
They have been considered as ``transitional'' nuclei, displaying a character that varies from vibrational to $\gamma$ unstable. Indeed,  detailed analyses (see Ref. \cite{kim,giannatiempo1998}) provided a good description of even-even Pd nuclei as pertaining to a region of transition from the vibrational U(5) limit to the $\gamma$-soft O(6) limit of the IBA-2 model \cite{iachello}. 

This interpretation has been recently questioned in a systematic study of the even mass isotopes of Mo, Ru, Pd, Cd, and Te \cite{garrett2018}. The authors concluded that the existence of low-energy quadrupole vibrations in some of these nuclei must be questioned and that the study of collective states must involve not only electromagnetic observable such as B(E2) values and quadrupole moments, which by definition only sample the charge and/or current distributions, but also other electromagnetic probes that are sensitive to shape coexistence and configuration mixing, such as, for instance, the electric monopole (E0) transitions. 

The question of whether Pd nuclei may actually exhibit a nearly-harmonic quadrupole structure has been recently addressed by two experiments involving the neutron inelastic scattering, devoted to the study of the structure of the $^{106}$Pd isotope \cite{prados,peters}. In the first one, a characterization of the low-lying excited states up to $\approx$2.4 MeV for spin $\leq6$ 
was obtained. The level scheme was organized into rotational bands, each characterized by a definite value of $K$. In the second experiment, on the basis of previously measured internal conversion electron  \cite{colvin} and new lifetime data, the strength of E0 transitions between 2$^+$ states were determined. The authors concluded that the extracted monopole transition strength values provide evidence for shape coexistence between the bands with the same $K$ value.

The existing data on conversion electrons for $^{106}$Pd isotope are rather limited and affected by a large uncertainty: for instance, two values differing by a factor $\approx$ 3 are available for the internal conversion coefficient of the $2_3^+\longrightarrow 2_1^+$ transition, thus preventing a definite conclusion on the amount of mixing between these two levels.



The aim of the present work is to provide further information to better understand the structure of low-lying levels in the Pd isotopes with N$\sim60$.
The E0 transitions between both 0$^+$ and $2^+$ states in $^{106}$Pd have been studied via internal conversion electron spectroscopy, performed by means of the apparatus that we have very recently developed \cite{marchini}. The new data, combined with those obtained in the re-analysis of data previously acquired on $^{104}$Pd by the nuclear spectroscopy group in Florence, help to clarify the properties of the $0^+$ and $2^+$ states up to 2.3 MeV in the $^{104,106}$Pd isotopes. 

\section{Experiment Details}
A dedicated experiment to study the structure of $^{106}$Pd at low excitation energy was performed at the INFN Legnaro
National Laboratories (LNL) in Italy. The nucleus of interest was populated in the EC-$\beta^+$ decay of $^{106g}$Ag (T$_{1/2}$=24~min) and $^{106m}$Ag (T$_{1/2}$=8~d) produced via the $(p,n)$ reaction on a self-supporting target of $^{106}$Pd 3~mg/cm$^2$ thick (96$\%$ enriched). The 5.5~MeV proton beam was delivered by the LNL Van der Graaff CN accelerator with an average intensity of 200~nA. In order to favor the fast decay activity of $^{106}$Ag, which populates the $0^+$ levels in $^{106}$Pd, measurements have been performed by alternating bombarding and measuring periods of 35~min. A 5~min waiting time was inserted to allow the decay of the short-lived $^{108}$Ag (T$_{1/2}$=3.4~min) produced in the $(p,n)$ reaction on the 1\% $^{108}$Pd isotope present in the target. The $^{108}$Ag beta decays mostly (95\%) to the ground state of  $^{108}$Cd ($Q(\beta^-)=1650(7)$ keV) increasing the background in the electron spectra. By inserting the above mention waiting time this background is reduced by a factor $\approx4$ while only the 10\% of the $^{106}$Ag activity is lost.

\begin{figure}
\includegraphics[width=\columnwidth]{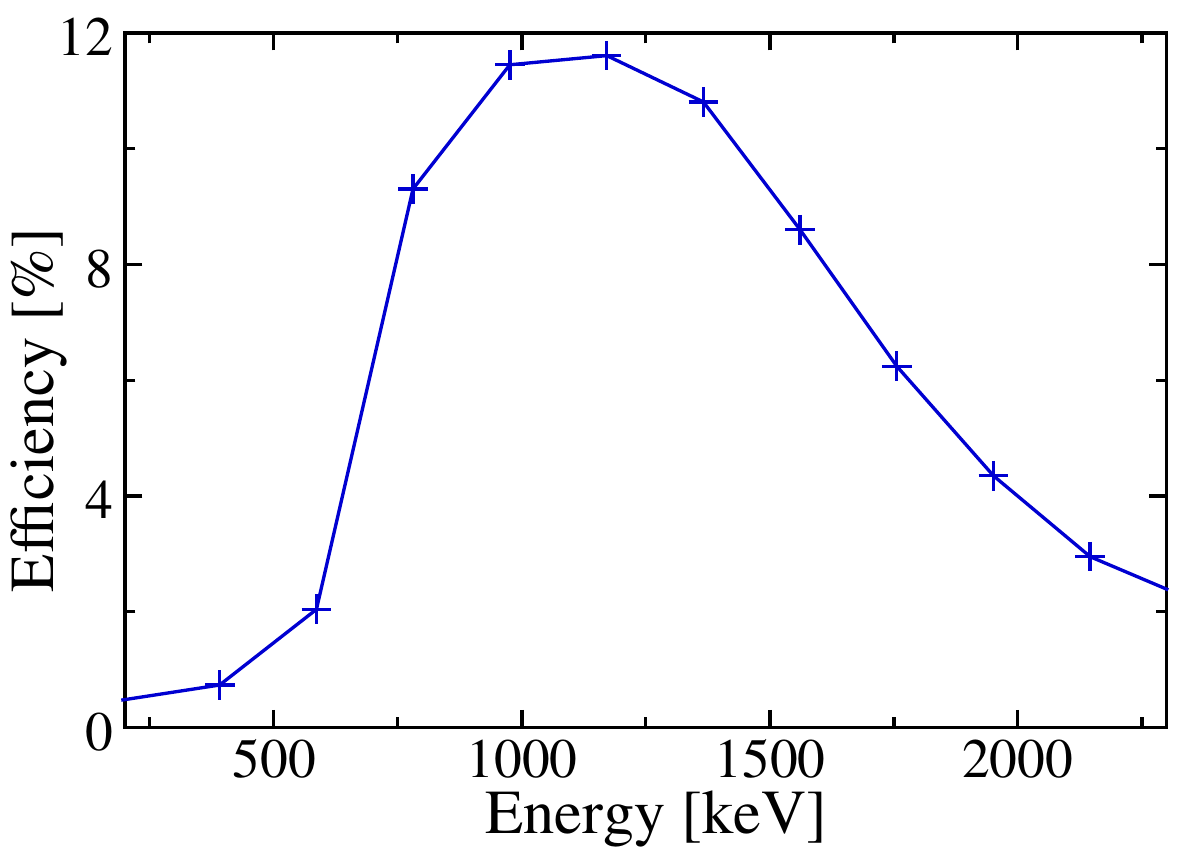}
\caption{GEANT4 simulated absolute efficiency curve of SLICES for the detector–source distance of 117 mm and using four magnet clusters. For more details, see Ref \cite{marchini}.}
\label{effi-mos}
\end{figure}

The internal conversion electrons emitted in the de-excitation of the states populated in the decay of $^{106g}$Ag were detected by the SLICES spectrometer \cite{marchini}, used for the first time in the present experiment. SLICES setup utilizes a 6.8 mm thick segmented lithium-drifted silicon detector coupled to a magnetic transport system to guide the electrons around a central photon shield towards the detector. The efficiency of the spectrometer can be optimized by changing the shape of the magnetic transport system components. For the configuration adopted in this experiment, the maximum of the efficiency curve is about 12$\%$ for transmitted energy of 1~MeV, as shown in Fig. \ref{effi-mos}. The adopted configuration and the related efficiency curve have been studied in detail in Ref. \cite{marchini}. 
An HPGe detector with an energy resolution of 2.4~keV (FWHM) at 1.3~MeV was used to detect $\gamma$ rays deexciting the nuclear states. 

\subsection{RESULTS}
\begin{table}
\begin{tabular}{ccccc}
\hline
$J_i^\pi \longrightarrow J_f^\pi$ & $E_{\gamma}$~[keV] & $\alpha_K^{exp.}\cdot10^3$ & $\alpha_K^{th}(E2)\cdot 10^3$ & $\alpha_{K}^{th}(M1)\cdot 10^3$ \\
\hline
$2_2^+ \longrightarrow 2_1^+$ & $616$  & $2.97(11)$ & $2.89$ & $2.97$ \\
$2_2^+ \longrightarrow 0_1^+$ & $1128$  & $0.64(9)$ & $0.68$ &  \\
$2_3^+ \longrightarrow 2_1^+$ & $1050$  & $1.06(7)$ & $0.79$ & $0.89$ \\
$0_2^+ \longrightarrow 2_1^+$ & $621$  & $2.6(2)$ & $2.8$ &  \\
$0_3^+ \longrightarrow 2_1^+$ & $1195$  & $0.71(13)$ & $0.60$ &  \\
$0_4^+ \longrightarrow 2_2^+$ & $873$  & $1.23(8)$ & $1.20$ & \\
\hline
\end{tabular}
\caption{Experimental K-internal conversion coefficients, $\alpha_K$, for transitions in $^{106}$Pd compared with the calculated values from BRICC \cite{bricc}.}
\label{alfa}
\end{table}

\begin{figure}
\includegraphics[width=\columnwidth]{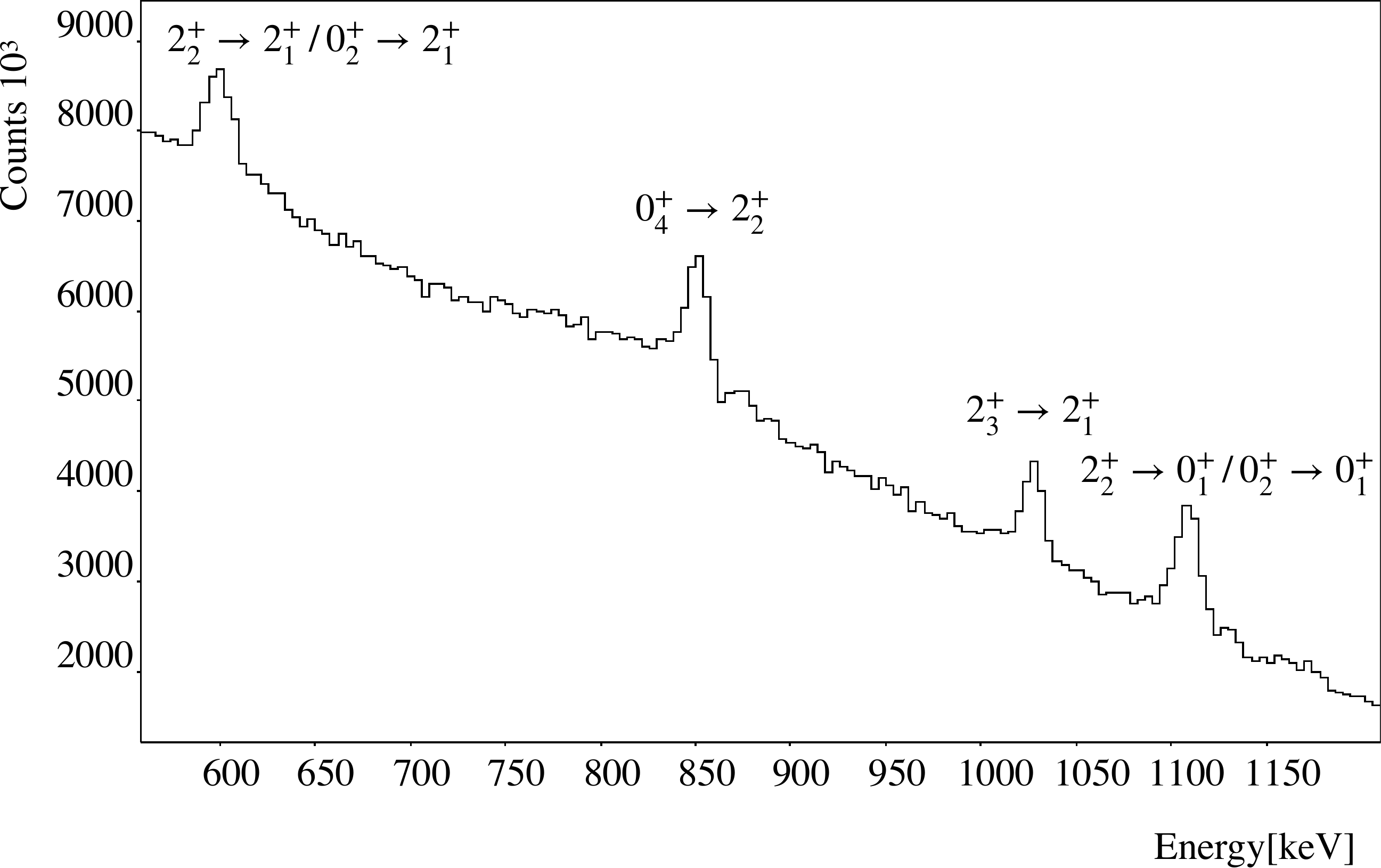}
\caption{Section of the SLICES energy spectrum; K-conversion lines are labeled. }
\label{Si}
\end{figure}

Conversion electron measurements have been performed to determine K-internal conversion coefficients, $\alpha_K$, and to evaluate the monopole strength, $\rho^2$(E0), of E0 transitions between states having the same spin and parity.

The $\alpha_K$ values obtained for the transitions of interest are summarized in Table \ref{alfa}. 
The agreement between the experimental and the theoretical values for pure E2 transitions on one hand is a test of the reliability of SLICES apparatus in performing in-beam measurements, on the other hand, of the correct determination of $\alpha_K(2_3^+ \longrightarrow 2_1^+)$.  
Two different $\alpha_K$ values for this transition are reported in the literature \cite{colvin,farzin}. The value obtained in the present work is in agreement with the one determined in Ref. \cite{colvin}. The experimental $\alpha_K(2_3^+ \longrightarrow 2_1^+)$ value, large with respect to the calculated one, suggests the presence of a strong E0 component in this transition.
The section of the electron spectrum in the energy range around 1~MeV is shown in Fig. \ref{Si}. The K-conversion electron peak of the $2_3^+ \longrightarrow 2_1^+$ transition is in a clean region of the spectrum. 

Measurements of internal conversion electrons can also provide information on the $\rho^2(E0$). For a transition between states with $J^+_i = J^+_f = 0$, it  is related to the ratio  \begin{equation}q^2_{ifj}={I_K(E0; 0_i^+\rightarrow 0_f^+)/ I_K(E2;0_i^+\rightarrow 2_j^+)}\label{q2}\end{equation} 
between the intensity of the E0 and E2 K-conversion lines de-exciting a given $0_i^+$ level.
The E0 strength can be determined via the expression:
\begin{equation}
    \rho^2(E0; J_i^+\rightarrow J_f^+) = q^2_{ijf}(E0/E2) \times \frac{\alpha_K(E2)}{\Omega_K(E0)} \times W_{\gamma}(E2)
\end{equation}
where $\Omega_K$ is the electronic factor for the K-conversion of the E0 transition obtained from Ref. \cite{bricc}, $\alpha_K$(E2) is the K-conversion coefficient for the E2 transition and W$_{\gamma}$(E2) is the $\gamma$-ray E2 transition probability. 

In the case of $J_i^+=J_f^+\neq0$, the E0 and E2 transitions in Eq. (\ref{q2}) connect the same initial and final levels. Since the contributions due to the different multipolarities to the same transition are indistinguishable, $q^2_{ijf}(E0/E2)$ is extracted from the internal conversion coefficient, which in the case of mixed E0, E2 and M1 multipolarities, has the expression: 


\begin{equation}
         \alpha_K={\alpha_K^{th}(M1)+(1+q_{ifj}^2)\cdot \delta^2 \cdot\alpha_K^{th}(E2) 
\over(1 + \delta^2)} 
    \end{equation}
where $\delta$ is the (E2/M1) mixing-ratio, and $\alpha_K^{th}(M1)$, $\alpha_K^{th}(E2)$ are the theoretical values of the internal conversion coefficient from the Band-Raman
Internal Conversion Coefficents (BRICC) database \cite{bricc}.

The $q^2$(E0/E2) and $\rho^2$(E0) values extracted in the present work are summarized in Table \ref{rho_exp}. 
The analysis of the $2_2^+ \longrightarrow 2_1^+$ K-electron line was made difficult by the presence of the predominant 616 keV peak due to the K-conversion electrons of the $0_2^+ \longrightarrow 2_1^+$ transition. As a consequence, the obtained $q^2$($2_2^+ \longrightarrow 2_1^+$) value has a large uncertainty.


\begin{table*}
\begin{tabular}{ccccccccc}
\hline
\multicolumn{5}{ c }{  } & \multicolumn{2}{c }{$q^2$(E0/E2)}& \multicolumn{2}{ c }{$\rho^2$(E0)$\cdot10^3$} \\
 \hline
$J_i^\pi \longrightarrow J_f^\pi$ & $E_{\gamma}$~[keV] & $\tau$~[fs] & $\delta(E2/M1)$ &  I$_\gamma$ & Present & Previous & Present & Previous \\
\hline
$0_2^+ \longrightarrow 0_1^+$ & $1134$&$8400(1900$)& & & $0.166(15)$ & $0.162(7)\footnote[1]{Reference \cite{smallcomb}.}$ & $17(4)$&$16.4(40)\footnotemark[1]$ \\
$0_3^+ \longrightarrow 0_1^+$ & $1706$  &$4000(700)$& &$ 0.857(34)$& $0.09(15)$ &  &$2(4)$& $<3\footnotemark[2]$ \\
$0_4^+ \longrightarrow 0_1^+$ & $2001$  &$>1200$& & & $0.124(18)$ & $ $ & $<19$&$ $ \\
$0_4^+ \longrightarrow 0_2^+$ & $867$  &$>1200$& & & $0.22(6)$ & $ $ & $<90$&$ $ \\
$2_2^+ \longrightarrow 2_1^+$ & $616$&$4500(360)$ &$-8.7^{+17}_{-19}$&$0.647(24)$&$0.027(38)$ & & $5(8)$&$ $ \\
$2_3^+ \longrightarrow 2_1^+$ & $1050$&$1900(190)$& $0.24(1)$& $0.853(34)$  & $4.2(18)$ & $5.8(33)\footnotemark[1]$&   $26(11)$ & $34(22)\footnote[2]{Reference \cite{peters}.}$\\
\hline
\end{tabular}
\caption{A comparison  between the E0 transition strengths $\rho^2$(E0) and $q^2$(E0/E2) extracted in the present work and in previous analyses. Transition energy, lifetimes for the parent state, multipole mixing ratios $\delta$(E2/M 1) and branching ratios I$_\gamma$ are taken from Ref. \cite{prados}}
\label{rho_exp}
\end{table*}

\begin{figure} \begin{center}
\includegraphics[width=1.4\columnwidth, angle=-90]{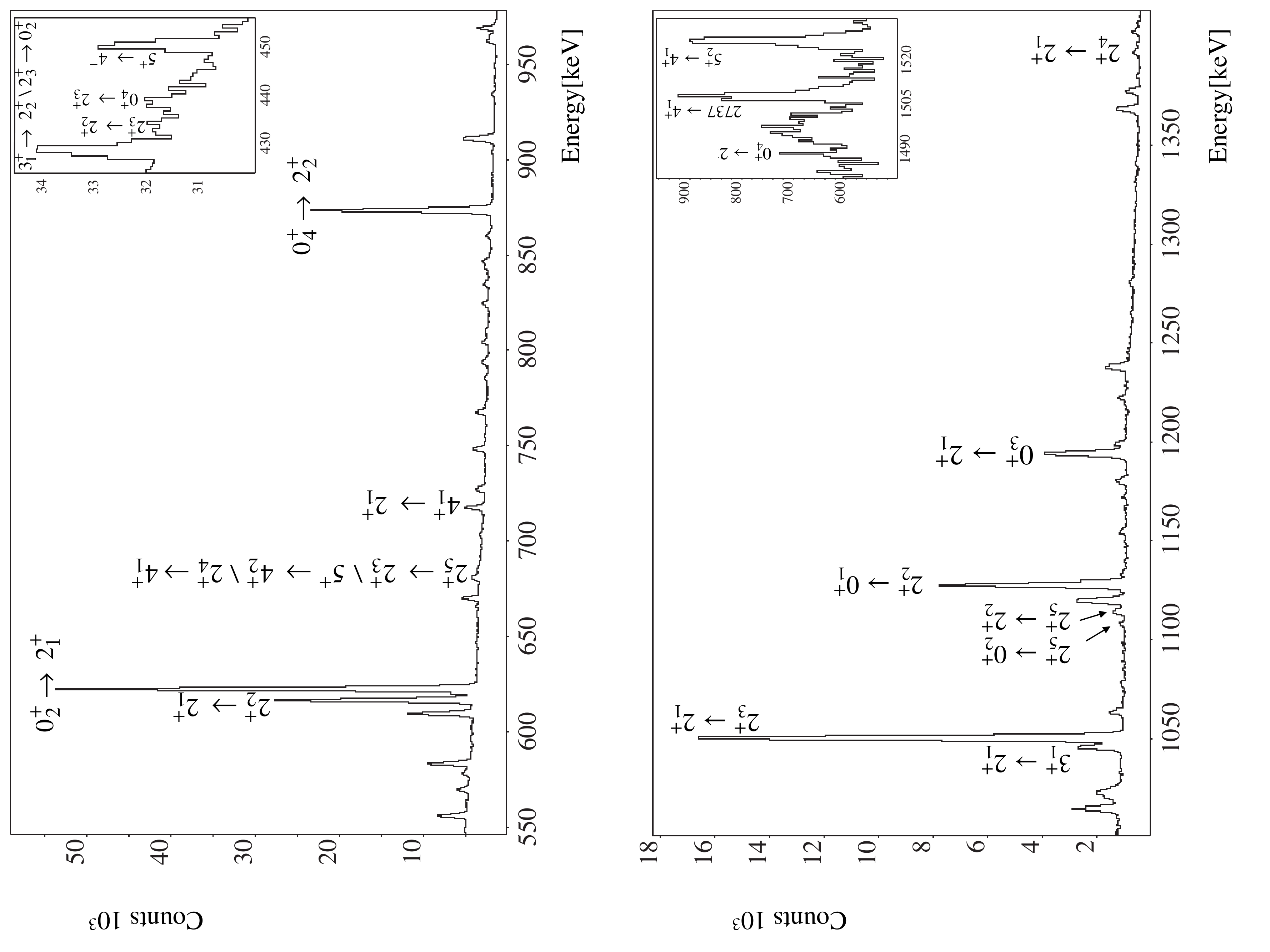}
\end{center}
\caption{Portions of the $\gamma$ ray energy spectrum. Some  peaks of interest are labeled with level spin and parity. The insets show the regions around 450 kev and 1500 keV respectively.}
\label{Ge}
\end{figure}
\begin{figure*}
\includegraphics[width=\textwidth]{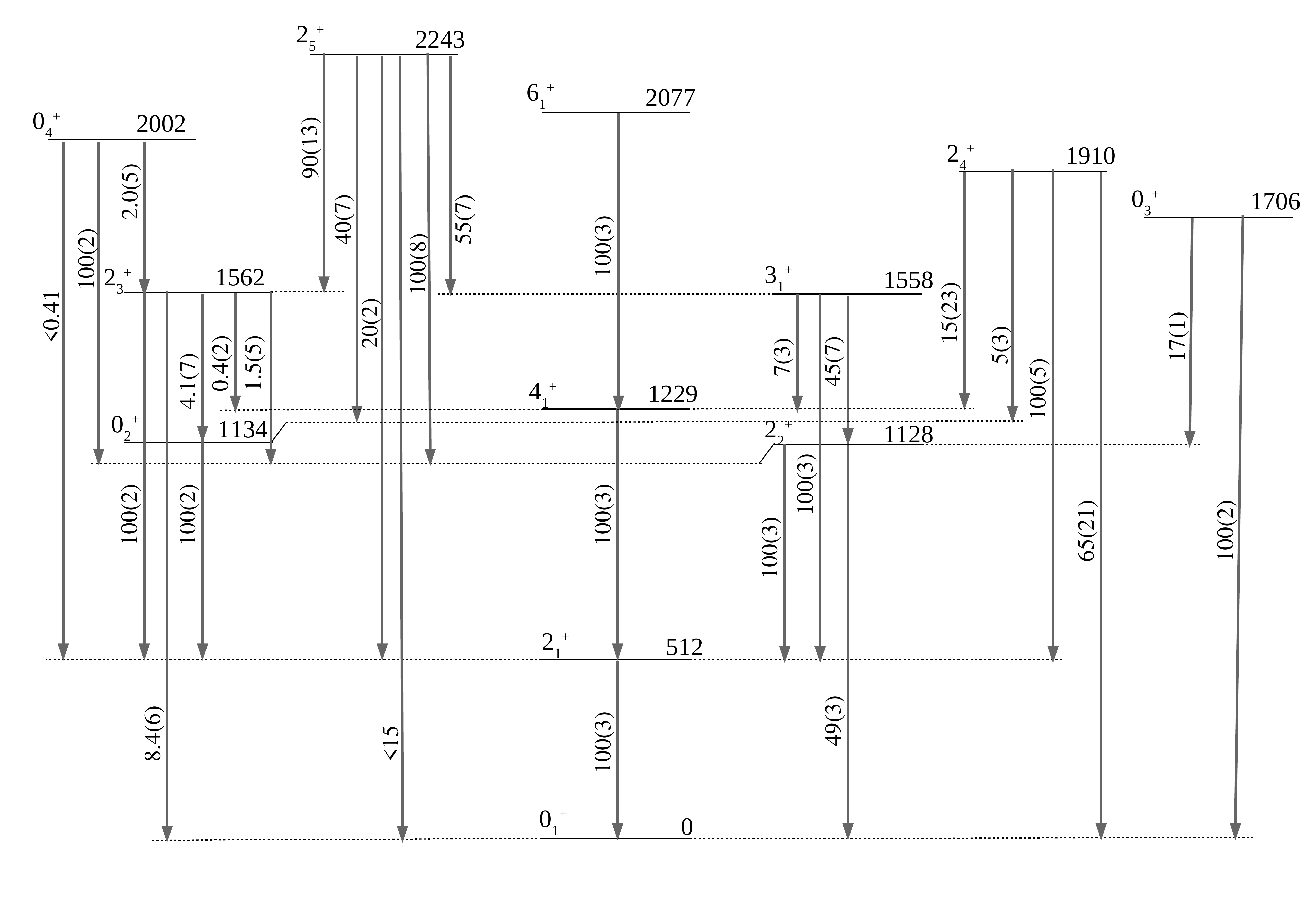}
\caption{Low-lying level scheme in $^{106}$Pd. The observed $\gamma$-transitions with the related branching ratios extracted in this experiment are reported on the arrow.}
\label{branching}
\end{figure*}
The coupling of SLICES with an HPGe detector allows us not only to extract the internal conversion coefficients but also to study in detail the decay scheme of the levels in $^{106}$Pd. An example of the $\gamma$ rays energy spectrum is reported in Fig. \ref{Ge}, where the transitions relevant to this work are shown. The part of the spectrum below 300 keV is dominated by the Compton edge of the 511 keV annihilation transition, which also covers the 512 keV, $2_1^+ \longrightarrow 0_1^+$ transition of $^{106}$Pd.
Fig. \ref{branching} shows the level scheme, up to an energy $\sim 2.3$ MeV. The decay branching ratios reported for each level were obtained in the present work. 
The level scheme of $^{106}$Pd has been recently studied also in a ($n,n'\gamma$) reaction \cite{prados}, where a number of new transitions are reported.  
We cannot confirm the existence of the 347~keV, $2_4^+ \longrightarrow 2_3^+$, 352~keV, $2_4^+ \longrightarrow 3_1^+$ and 782~keV, $2_4^+ \longrightarrow 2_2^+$ transitions since their intensities are below the sensitivity level of this work.
 As to the peak at 680~keV, it is too intense to be due only to the $2_5^+ \longrightarrow 2_3^+$ and $5_2^+ \longrightarrow 4_3^+$ well-known transitions. This could provide a hint of a possible contribution to this peak due to the new proposed $2_4^+ \longrightarrow 4_1^+$ transition.
 A 439 keV transition from the $0_4^+$ state to the $2_3^+$ one, not reported in \cite{prados}, is visible in the $\gamma$-spectrum (see inset in Fig. \ref{Ge} upper panel).  A small peak at an energy of $\sim 1489$ keV has been assigned to the $0_4^+ \longrightarrow 2_1^+$ transition (see inset in Fig. \ref{Ge} lower panel). Both these transitions were previously reported in Ref. \cite{nndc}.

Some years ago the Florence spectroscopy group has performed measurements of internal conversion electrons to investigate E0 transitions in $^{104}$Pd. The deduced E0 strengths are reported in Table III. In the same experiment, $\gamma-\gamma$ coincidences have been also measured \cite{bellizzi}.  
We have now re-analyzed the data in order to gain a deeper insight on the existence of a $0^+_4$ state, reported in Ref. \cite{nndc} at 2103(2) keV. This level has been seen only in a ($p,p'$) reaction and no information is given on its decay properties. 
In the $\gamma$ spectrum acquired in coincidence with the 786 keV $2_2^+\longrightarrow2_1^+$ transition, a small peak is visible (see. Fig. 5) at 759.3(5)~keV. In the single spectra, this peak is completely dominated by the more intense 758.8 keV  $4_2^+\longrightarrow4_1^+$ transition.  
\begin{figure}
\begin{center}
\includegraphics[width=\columnwidth]{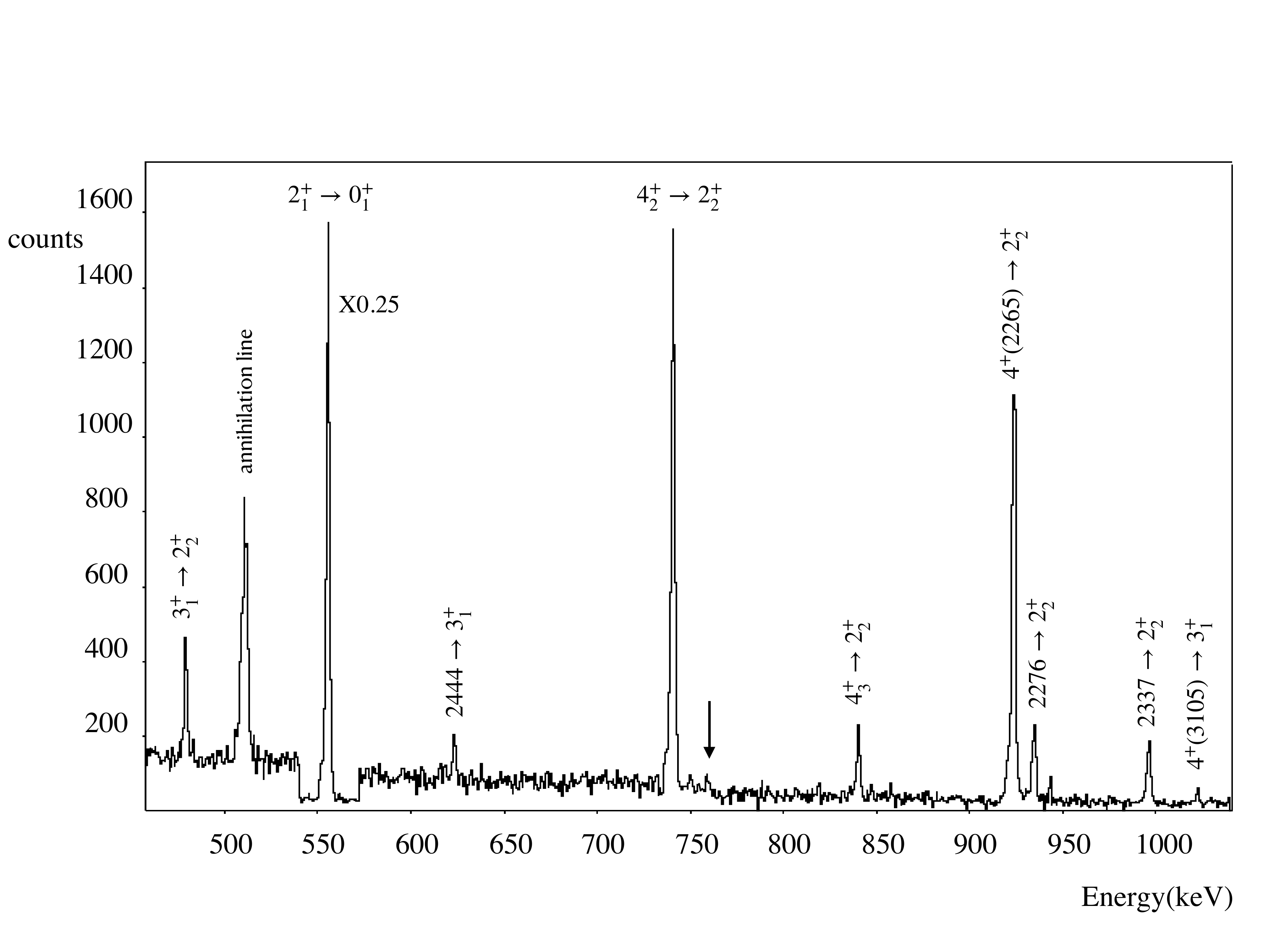}
\end{center}
\caption{Section of the $\gamma$ spectrum gated on the 786 keV, $2_2^+\longrightarrow2_1^+$  transition showing the region around 750 keV. A small peak is visible at an energy of 759 keV (indicated by the arrow).}
\label{gate786}
\end{figure}
Assuming that the peak corresponds to the $0_4^+\longrightarrow 2_2^+$ transition, the energy of the initial level would be 2101.0(5) keV. We looked for possible decays from this level to the $2_1^+$ state and to the $0_2^+$, $0_3^+$ ones in the $\gamma$ and electron spectra, respectively. The energies corresponding to the $\gamma$ and E0 transitions would be 1545.2 keV, 743.0(5) keV, and 283.8(5) keV, respectively. A new 1545.2(3) keV transition was indeed identified in Ref. \cite{bellizzi} but was assigned to the decay from the 2868.7 keV level. 
The 743 keV E0 transition in our data would be completely covered by the much more intense K-conversion line of the 768 keV, $4_1^+\longrightarrow 2_1^+$ transition, while a small peak at 284 keV is visible in the electron spectrum of Fig. \ref{ele104pd}.  Since in the corresponding $\gamma$ spectrum (Fig. \ref{ele104pd}, lower panel), there is no peak at an energy $\sim 308$ keV (while the peak corresponding to the 289 keV transition from the $4^+$ level at 3158 keV is clearly visible) we tentatively assign E0 multipolarity to the transition and hence spin-parity
$0^+$ to the level at an energy of 2101 keV in $^{104}$Pd.
\begin{figure}
\begin{center}
\includegraphics[width=1.4\columnwidth, angle=-90]{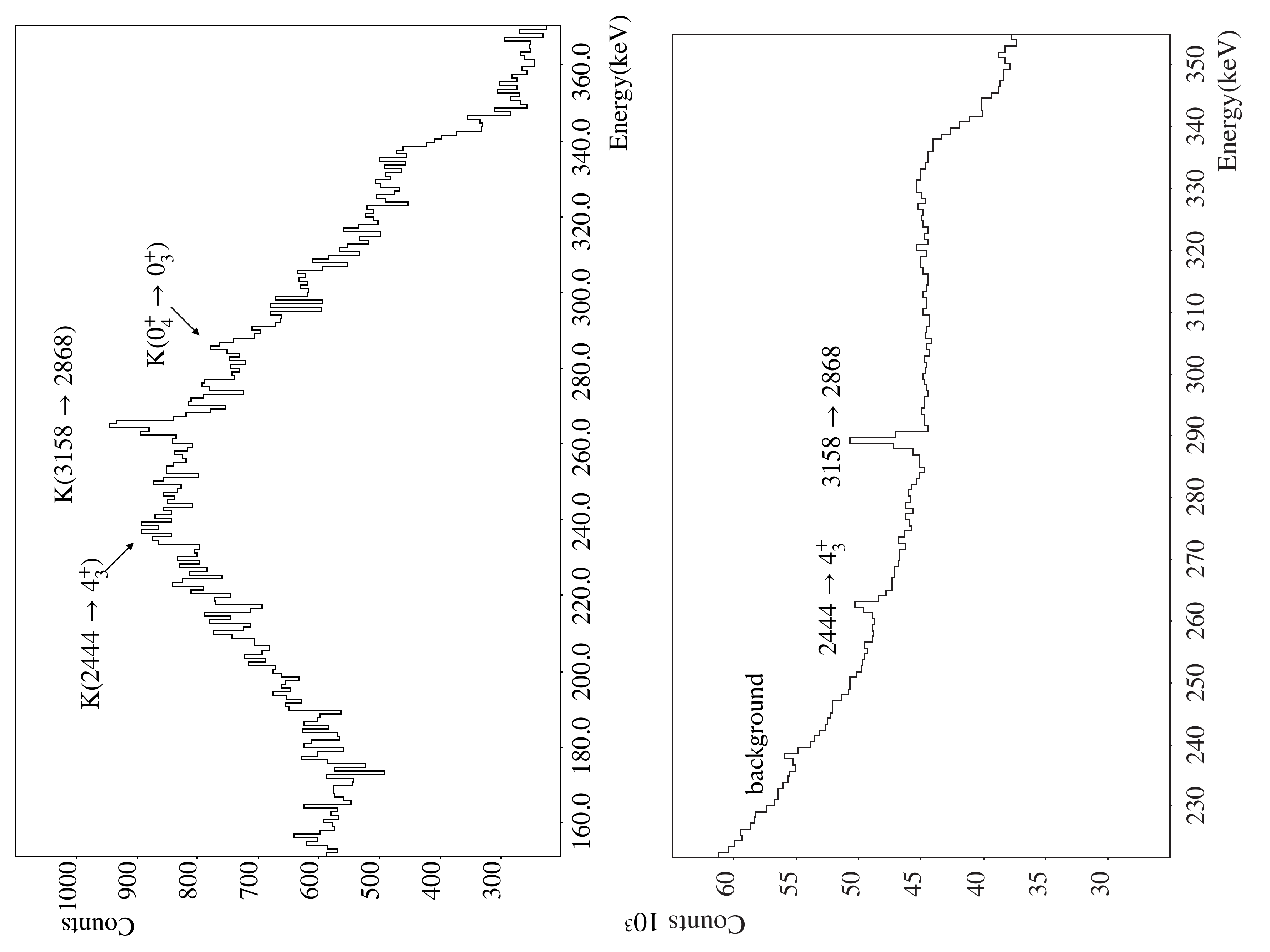}
\end{center}
\caption{Upper panel: Section of the electron spectrum in the 230 -- 300 keV energy range. The small peak visible at an energy of 284 keV has been assigned to the $0_4^+\longrightarrow 0_3^+$ transition. Lower panel: Portion of the $\gamma$ spectrum in the 250 -- 350 energy range. Only the peak corresponding to the 289 keV transition from the $4^+$ level at 3158 keV is visible.}
\label{ele104pd}
\end{figure}
 
\section{Discussion}

\begin{figure*}
\includegraphics[width=\textwidth]{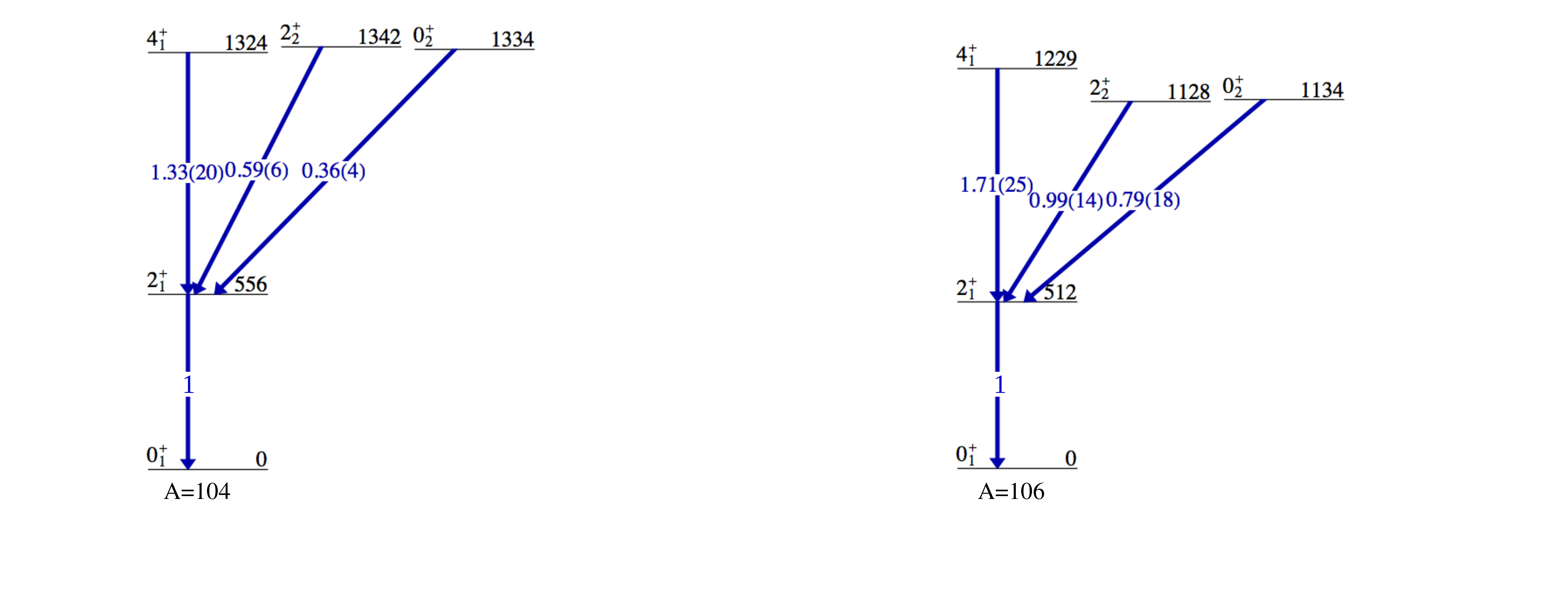}
\caption{Low-lying levels in even-even $^{104,106}$Pd isotopes. 
The B(E2) transition strengths normalized to the B(E2; 2$_1^+ \rightarrow$ 0$_1^+$) value are reported on the arrows. Data are taken from Ref. \cite{nndc}.}
\label{multiphonon}
\end{figure*}

The interpretation of low-lying levels in even mass Pd isotopes is still controversial and different models have been employed to describe their properties.
In the present work, we consider the contribution that the study of E0 transition can provide to clarify the structure of the low lying states in $^{104-106}$Pd.

The excitation energy pattern of the low-lying levels in the $^{104-106}$Pd isotopes might suggest a vibrational structure, with a triplet of states with $J^\pi=0^+,\,2^+,\,4^+$ whose energy is approximately twice that of the first $2^+$ state. However, the B(E2) values of transitions from these states to the $2_1^+$ cast some doubts on their vibrational character. 
The values of the ${B(E2; I^\pi \rightarrow 2_1^+)}$ for the decays of the two-phonon states ($0^+$,$2^+$,$4^+$) should be identical and twice the value of the ${B(E2; 2^+_1 \rightarrow 0^+_1)}$ one-phonon decay, instead they differ considerably and are smaller than expected (see Fig. \ref{multiphonon}).

As to the identification of the three-phonon quintuplet, it is made difficult by the presence of additional levels with $J^\pi=0^+,\,2^+$. They have been considered as intruder states resulting from proton-pair excitations across the Z = 50 shell. Two signatures are commonly given for the identification of an intruder states: i) the characteristic V-shape pattern of their excitation energies versus neutron number ii)  the enhanced cross-section for single- and two-nucleon transfer reactions with respect to those between collective states. Low-lying intruder configurations have been studied in even-even Pd isotopes in Refs. \cite{Lhersonneau,wang}. Based on the energy systematic, the 0$^+_3, 2_4^+$ pair of states is suggested to have intruder character until N=60 and again for N=70, while the pair $0_2^+, 2_3^+$ becomes intruder state for N=62,64. Within this hypothesis, the V-shaped pattern of the excitation energy is granted.  The interpretation of the $0_3^+$ states as intruder states in $^{104,106}$Pd isotopes was also supported in Ref.\cite{giannatiempo1998} by the analysis of their decay properties.
The only available data for the ($^3$He,$n$) transfer reaction is an upper limit for the cross-section to the 0$^+_2$ for N=58 isotope reported in Ref. \cite{garrett2016}, which is much smaller than the ground-state to ground-state cross-section in $^{104}$Pd and $^{106}$Pd. 
 
A detailed analysis of excitation energy patterns and electromagnetic properties of positive-parity levels in even $^{100-116}$Pd (Z = 46) was performed some years ago \cite{giannatiempo1998} in the framework of the IBA-2 model, which is particularly suitable to study the evolution of an isotopic chain as a function of the neutron number. In that work, all the excitation energies and electromagnetic properties, available at the time for the low-lying levels in the even $^{100-116}$Pd isotopes,  were investigated, with the exception of E0 transitions.
 An analogous study had been also performed by the same authors in the even $^{98-114}$Ru isotopes \cite{giannatiempoRu}. The parameters were requested to vary smoothly along each isotopic chain and among isotones in neighboring isotopic chains. An overall satisfactory agreement was obtained. The conclusion was that the even palladium isotopes could be considered as lying close to a transitional region between the vibrational U(5) limit to the $\gamma$-soft O(6) limit of the IBA-2 model.

\begin{figure}
\includegraphics[width=\columnwidth]{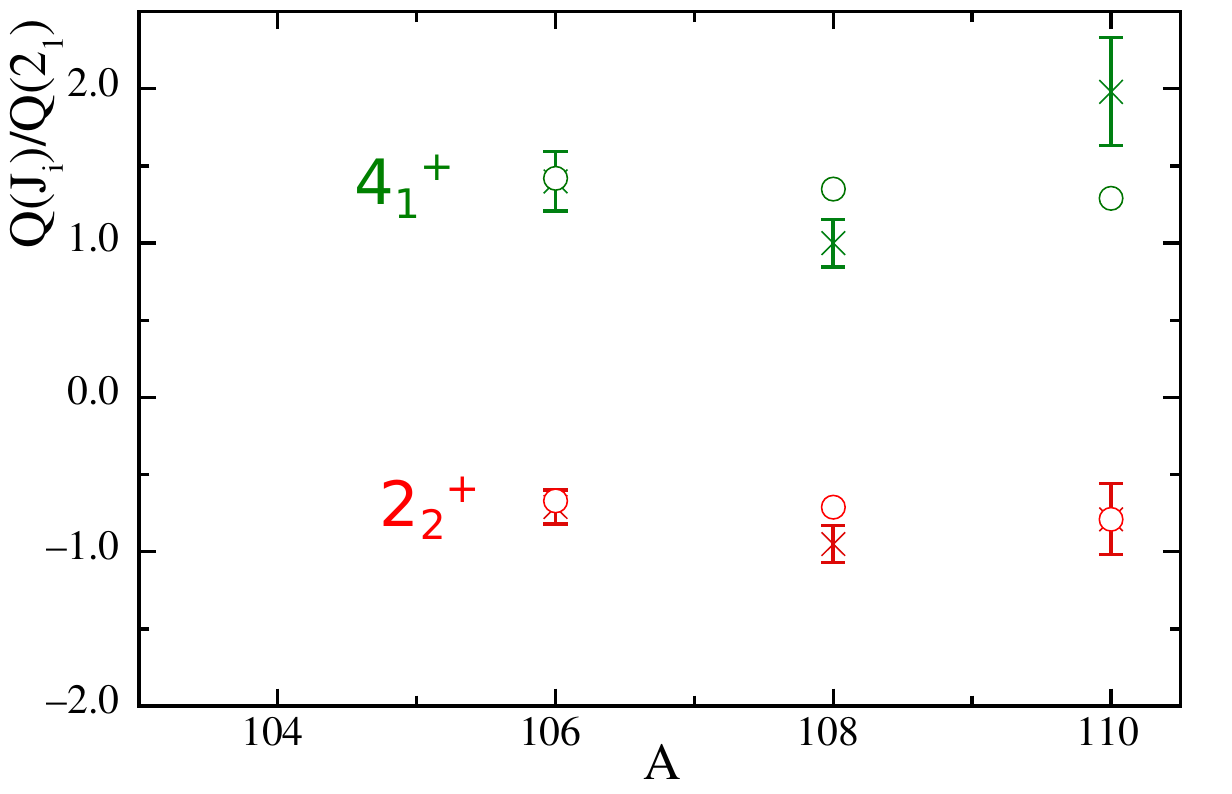}
\caption{Experimental values (crosses) of Q for the $2_2, 4_1$ two-phonon candidates, normalized to the Q(${2_1}$) as a function of mass number. Data for the 4$_1^+$ and 2$_2^+$ levels are reported in green and red respectively. The values are taken from Ref. \cite{giannatiempo2018}. }
\label{Q_values}
\end{figure}

More recently a new IBA-2 work has been published \cite{giannatiempo2018}, which uses the parameters of Ref. \cite{giannatiempo1998} to study the large body of new experimental data become available over the years on the even Pd isotopic chain. In this analysis, which is mainly centered on the vibrational-$\gamma$ band structure, also all the known quadrupole moments were taken into account.  
A comparison between the available experimental values (from Ref. \cite{nndc}) and the calculated values from Ref. \cite{giannatiempo2018} of the $Q$ normalized to the $Q(2_1^+)$ for the 2$^+_2$ and 4$^+_1$ levels is reported in Fig. \ref{Q_values}. The agreement is good and the calculations are able to correctly predict the inversion of sign measured for the $Q(2_2^+)$.

An interpretation very different from that of Ref. \cite{giannatiempo1998, giannatiempo2018} was given in  Ref. \cite{garrett2018}. 
Here, the authors compare the properties of the low-lying levels of $^{102-110}$Pd with the predictions of the harmonic vibrator and it is underlined that in none of the considered Pd isotopes the B(E2) values for decays to the $2_1^+ $ state meet the vibrational requirements. The conclusion is that the harmonic spherical vibrator interpretation breaks down already at the two-phonon levels.  In particular, in the $^{106}$Pd isotope the author assigns the $0_2^+$ state as the head of an intruder shape-coexisting band (in agreement with Ref. \cite{prados}) while the $2_5^+$ is suggested to be the member of a $\gamma$ band built on the $0_2^+$ state.

In order to further clarify to what extent the interpretation of the Pd isotopes in the framework of the IBA-2 model is valid, we performed an analysis of the available experimental data on the E0 transition between the low-lying states in $^{104-106}$Pd. 
The analysis has been performed by using the Hamiltonian and the parameters of Ref. \cite{giannatiempo1998}. The Hamiltonian has been diagonalized in the U$_{\pi,\nu}(5)$ basis, using the NPBOS code \cite{npbos}, which gives in its output the d-boson number components for each state. 
Excitation energies, E2 and M1 transitions in $^{104,106}$Pd isotopes have been already investigated in detail in Ref. \cite{giannatiempo1998}. In the present work, we limited the analysis to the monopole strengths between low-lying 0$^+$ and 2$^+$ levels, which were not previously considered. 

In the IBA-2 model the E0 transition operator has the expression \cite{iachello}:
\begin{equation}\begin{split}
    \hat{T}(E0)&=\beta_{0\nu}\hat{T}_\nu(E0)+\beta_{0\pi}\hat{T}_\pi(E0) \\
    &= \beta_{0\nu}(d^\dagger_\nu \times \tilde{d}_\nu)^{(0)} + \beta_{0\pi}(d^\dagger_\pi \times \tilde{d}_\pi)^{(0)}
 \end{split}    \end{equation}
\begin{equation} \begin{split}
    \rho^2(E0; J_i^+\rightarrow J_f^+)=\frac{Z^2}{e^2R^4}&[\beta_{0\nu} \langle J_f | \hat{T}_\nu(E0) | J_i \rangle  \\
    &+\beta_{0\pi} \langle J_f | \hat{T}_\pi(E0) | J_i \rangle]^2
\end{split} \end{equation}
where R=1.2A$^{1/3}$~fm, and the parameters $\beta_{0\nu}$ and $\beta_{0\pi}$ are expressed in $e$ fm$^2$.

One of the biggest difficulties in the study of the E0 transitions is related to the lack of systematics on the values of the E0 effective charges, which prevents defining a range of proper values. In the present work, in order to evaluate the effective monopole charges, the experimental data on $\rho^2$(E0) have been compared with the corresponding theoretical values by performing a standard $\chi^2$ minimization procedure restricted to the range [-1,+1] $e$ fm$^2$. The $\rho^2$(E0) values used in the comparison are marked with asterisks in Table \ref{E0_calc}. We included the $\rho^2(E0; 0_2^+\longrightarrow0_1^+)$ values measured in the isotone $^{100-102}$Ru nuclei to further constrain the minimization procedure. 

\begin{figure} \begin{center}
\includegraphics[width=\columnwidth]{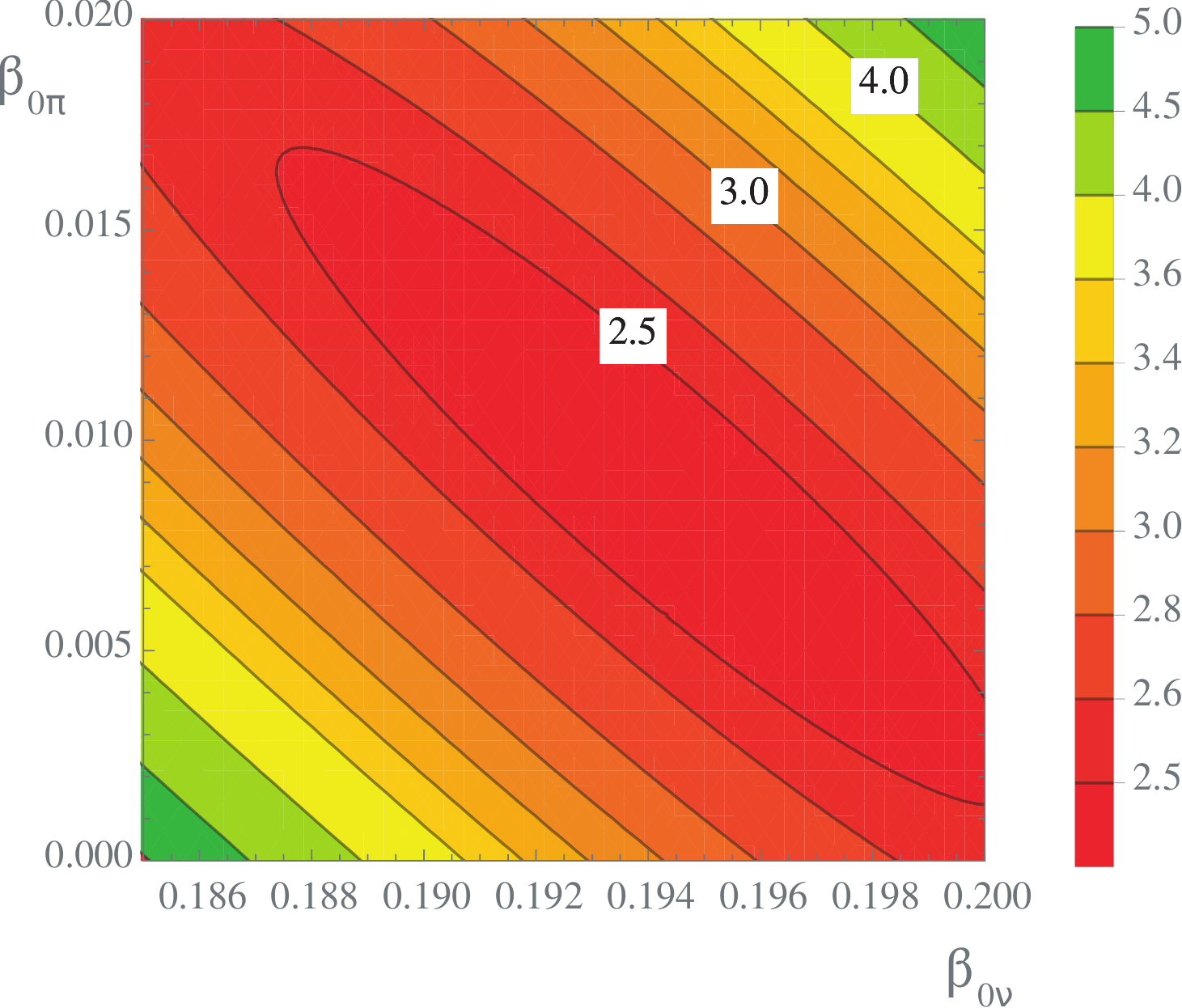}\end{center}
\caption{Contour plot for the reduced $\chi^2$ variable based on the comparison of theoretical and experimental $\rho^2$(E0) values as a function of the effective monopole charges  $\beta_{0\nu}$ and $\beta_{0\pi}$ (in $e$~fm$^2$). }
\label{chi2}
\end{figure}

In the comparison we assumed that the $0^+$ intruder state is the $0_3^+$ level in $^{104,106}$Pd.
In Fig. \ref{chi2} the contour plot for the normalized $\chi^2$ is reported. The minimum is centered at $\beta_{0\nu}$=0.194~$e$~fm$^2$ and $\beta_{0\pi}$=0.009~$e$~fm$^2$.

By using these values for the effective monopole charges we have calculated the $\rho^2$(E0)  reported in Table \ref{E0_calc}.  

\begin{table} 
\begin{tabular}{cccccc}
\hline
Nuclide&$J_i^\pi \longrightarrow J_f^\pi$ & $E_{\gamma}$~[keV] & $\rho^2(E0)_{exp}\cdot10^3$ & & $\rho^2(E0)_{calc}\cdot10^3$ \\
\hline
$^{104}$Pd & $0_2^+ \longrightarrow 0_1^+$ & $1334$  &$11(2)\footnote[1]{Reference \cite{bellizzi}.}$&*& $10$  \\
$^{104}$Pd & $2_2^+ \longrightarrow 2_1^+$ & $786$  &$5(4)\footnote[2]{Calculated in the present work from the data of Ref. \cite{bellizzi}.}$&*& $1$  \\
\hline
$^{106}$Pd & $0_2^+ \longrightarrow 0_1^+$ & $1134$  &$17(4)\footnote[3]{Present work.}$&*& $16$  \\
$^{106}$Pd & $0_4^+ \longrightarrow 0_1^+$ & $2001$   &$<19\footnotemark[3]$& & $0.3$  \\
$^{106}$Pd &  $0_4^+ \longrightarrow 0_2^+$ & $867$   &$<90\footnotemark[3]$& & $4$  \\
$^{106}$Pd & $2_2^+ \longrightarrow 2_1^+$ & $616$  &$5(8)\footnotemark[3]$ & & $1$   \\
$^{106}$Pd & $2_3^+ \longrightarrow 2_1^+$ & $1050$  & $26(11)\footnotemark[3]$&*& $28$   \\
$^{106}$Pd & $2_4^+ \longrightarrow 2_1^+$ & $1398$ &$21^{+10}_{-21}\footnote[4]{Reference \cite{smallcomb}.}$& & $0.1$\\
  &  &   &$18^{+10}_{-18}\footnotemark[4]$& \\
$^{106}$Pd & $2_5^+ \longrightarrow 2_2^+$ & $1115$ &$96^{+43}_{-61}\footnotemark[4]$ & & $18$ \\
\hline
$^{100}$Ru & $0_2^+ \longrightarrow 0_1^+$ & $1130$  &$10.3(18)\footnote[5]{Reference \cite{kibedi}}$&*& $11.4$  \\
$^{102}$Ru & $0_2^+ \longrightarrow 0_1^+$ & $944$  &$14(3)\footnotemark[5]$&*& $17$  \\
\hline
\end{tabular}
\caption{Experimental values of $\rho^2$(E0) in $^{104,106}$Pd and $^{100,102}$Ru compared to theoretical ones evaluated using the Hamiltonian parameters from Ref. \cite{giannatiempo1998} and the E0 effective charges $\beta_{0\nu}$=0.194~$e$~fm$^2$, $\beta_{0\pi}$=0.009~$e$~fm$^2$ deduced in the present work. The values marked by an asterisk have been used in the $\chi^2$ minimization procedure.}
\label{E0_calc}
\end{table}
Limiting our considerations to the Pd isotopes, we note the agreement between experimental and calculated values of the $\rho^2$(E0) for the transition de-exciting the $0_2^+$ levels supporting the interpretation of this state as belonging to the IBA-2 model space. Also for the $2_2^+$ level, we observe that the IBA-2 calculations of the $\rho^2$(E0) values do not contradict the interpretation of these states as belonging to the $n_d=2$ triplet.
For what concerns the fourth experimental 0$^+$ state in $^{106}$Pd, the calculated $\rho^2$(E0; $0_4^+ \longrightarrow 0_1^+$) value is much smaller than the $\rho^2$(E0; $0_4^+ \longrightarrow 0_2^+$) one as suggested by the experimental limits. 
A comparison between the experimental B(E2) values and the calculated ones for this state is reported in Table \ref{E2_calc}. The experimental values for the $0_4^+$ state are calculated using the limit on the lifetime recently reported in Ref. \cite{prados} and the branching ratios from the present analysis. It preferentially decays to the 2$^+_2$ state as expected for the member of the $n_d=3$ quintuplet. For $^{104}$Pd, in the present work hints for the existence of the fourth experimental 0$^+$ state at 2101~keV which preferentially decays to the 2$^+_2$ state are presented, but no experimental B(E2) values from this level are known.

The agreement found between the calculated and experimental $\rho^2$(E0; $2_3^+ \longrightarrow 2_1^+$) values seems to exclude the interpretation of this state as a member of an intruder band. 
Since also all the other electromagnetic properties of this state were reasonably reproduced by the calculations in Ref. \cite{giannatiempo1998} we are led to confirm that it lies within the model space.
We note that the IBA-2 calculations closely reproduce the experimental value of the B(E2; $2_3^+ \longrightarrow 4_1^+$), which is not included in the decay scheme proposed in Ref. \cite{prados,garrett2018}.

The scenario is different for the $\rho^2$(E0; $2_4^+ \longrightarrow 2_1^+$) and $\rho^2$(E0; $2_5^+ \longrightarrow 2_1^+$) strengths still known with limited precision. In Table \ref{E2_calc} the comparison between the experimental B(E2) and B(M1) values (from Ref. \cite{prados}) and the calculated ones of the 2$^+_4$ and 2$^+_5$ states is reported. 
For both states, the electromagnetic transition probabilities are known with a large uncertainty so that the comparison with the calculation is not decisive. As a consequence, no definite conclusion could be drawn on the interpretation of the $2_4^+$ as a member of the intruder band built on the 0$^+_3$ state, as suggested in Ref. \cite{wang}. We note also that no evidence of the $2^+_{4} \rightarrow 0^+_{3}$ transition has been reported so far and also in the present work no such transition was observed. Similarly, no definite conclusions can be drawn on the character of the state $2_5^+$. However since this state does not decay to the  $4_1^+$, while the $2_4^+$ does, it seems preferable to associate the $2_4^+$ to the $2^+$ member of a $n_d=3$ quintuplet and the $2_5^+$ level to a coexisting configuration as suggested in Ref. \cite{garrett2018}.

\begin{table*} 
\begin{tabular}{cccccc}
\hline
$J_i^\pi \longrightarrow J_f^\pi$ & $E_{\gamma}$~[keV] & $B(E2)_{exp}$ & $B(E2)_{calc}$ & $B(M1)_{exp}$ & $B(M1)_{calc}$ \\
\hline
 $0_4^+ \longrightarrow 2_3^+$ & $439$  &$<32$& $30$ \\
 $0_4^+ \longrightarrow 2_2^+$ & $873$  &$<1600$& $1500$  \\
 $0_4^+ \longrightarrow 2_1^+$ & $1489$  &$<2$& $3$  \\
 $2_4^+ \longrightarrow 2_3^+$ & $347$  &$3600^{+2400}_{-3600}$ & $14$   \\
  $2_4^+ \longrightarrow 4_1^+$ & $680$ &$155^{+9}_{-8}$ & $240$ \\
 $2_4^+ \longrightarrow 0_2^+$ & $776$  & $60^{+36}_{-30}$& $515$   \\
 $2_4^+ \longrightarrow 2_2^+$ & $782$  & $13^{+12}_{-8}$& $230$& $0.13^{+0.29}_{-0.10}$ & $7 \cdot 10^{-4}$  \\
 $2_4^+ \longrightarrow 2_1^+$ & $1397$ &$0.4^{+0.4}_{-0.2}$&  $2$& $7.3^{+4.2}_{-3.8}$ & $5.1$\\
  & &$36^{+23}_{-20}$&  $2$& $2.8^{+2.2}_{-1.6}$ & $5.1$\\
 $2_4^+ \longrightarrow 0_1^+$ & $1909$ &$5.7^{+3.3}_{-3.0}$ & $0.4$ \\
 $2_5^+ \longrightarrow 2_3^+$ & $680$ &$1400^{+900}_{-700}$ & $680$ & $42^{+43}_{-32}$ & $0.543$  \\
  $2_5^+ \longrightarrow 3_1^+$ &$684$ &$1600^{+690}_{-600}$ &  $44$ &$3.8^{+3.8}_{-2.0}$&$0.10$\\
 $2_5^+ \longrightarrow 0_2^+$ & $1109$ &$160^{+6}_{-7}$ & $1$ \\
 $2_5^+ \longrightarrow 2_2^+$ & $115$ &$100^{+150}_{-70}$ & $1.7$ & $19^{+15}_{-11}$ & $0.87$  \\ 
 $2_5^+ \longrightarrow 2_1^+$ & $1731$ &$5.0^{+25}_{-21}$ & $1.2$ & $0.177^{+22}_{-11} $ & $0.7$ \\ 
 &  &$(6^{+155}_{-6}) \cdot 10^{-3}$ & $1.2$ & $(1.20^{+47}_{-42}) \cdot 10^{-3}$ & $(6.97) \cdot 10^{-4}$ \\  
 $2_5^+ \longrightarrow 0_1^+$ & $2242$ &$1.8^{+7}_{-6}$ & $1.2$ \\ 
 \hline
\end{tabular}
\caption{The experimental values of B(E2) in $10^{-4}\,e^2b^2$ and B(M1) value in $10^{-3}\,\mu_N^2$ of the transition de-exiciting the $0^+_4$, $2^+_4$ and $2^+_5$ levels of $^{106}$Pd. The experimental data are taken from Ref. \cite{prados}. The theoretical values have been evaluated using the parameters from Ref. \cite{giannatiempo1998}. }
\label{E2_calc}
\end{table*}

The presence of a large transition strength is considered as a signature of strong mixing between two states with different deformation (Ref. \cite{wood}) and 
according to Ref. \cite{peters} the $\rho^2(E0)$ values measured in  $^{106}$Pd are large enough to provide evidence for shape coexistence in this nucleus.
We have therefore compared the experimental value of  $\rho^2(0_i^+ \longrightarrow 0_f^+)$ with that evaluated in a simple mixing model, following the procedure described in Refs. \cite{wood,giannatiempo_kr}.  The $0^+_2$ and $0^+_1$ states are assumed to be a linear combination of two basic configurations $\vert1\rangle$ and $\vert2\rangle$ of different deformations:
\begin{equation}
\vert0_1^+\rangle=b\vert1\rangle+a \vert2\rangle\,\,\,\,\,\, \vert0_2^+\rangle=a\vert1\rangle-b \vert2\rangle\,\,\,\,\, (a^2+b^2=1)
\end{equation}
It is possible to deduce an approximate expression for the monopole operator in terms of the deformation variables in a quadrupole deformation space \cite{davynov}:
\begin{equation}
\hat{T(E0)}={3Z\over4\pi} \left( \beta^2+{5\sqrt5\over 21\sqrt\pi} \beta^3 cos3\gamma \right)
\label{rozeroth}
\end{equation} 
In this approximation one obtains for $\rho^2(0_2^+ \longrightarrow 0_1^+)$ the expression:
\begin{equation}\begin{split}
\rho^2(0_2^+ \to 0_1^+)=&({3Z\over4\pi})^2a^2(1-a^2)[(\beta_1^2 -\beta_2^2) \\
&+{5\sqrt5\over 21\sqrt\pi}(\beta_1^3 cos3\gamma_1 -\beta_2^3 cos3\gamma_2)]^2
\label{rozeroth}
\end{split}
\end{equation} 
by neglecting the  non-diagonal  term $\langle2 \vert T(E0) \vert1\rangle$. The parameters $\beta_1,\,\,\gamma_1$  and $,\beta_2,\,\,\gamma_2$  refer to the $\vert1\rangle$ and $\vert2\rangle$ unmixed states, respectively.
\begin{figure}
\includegraphics[width=\columnwidth]{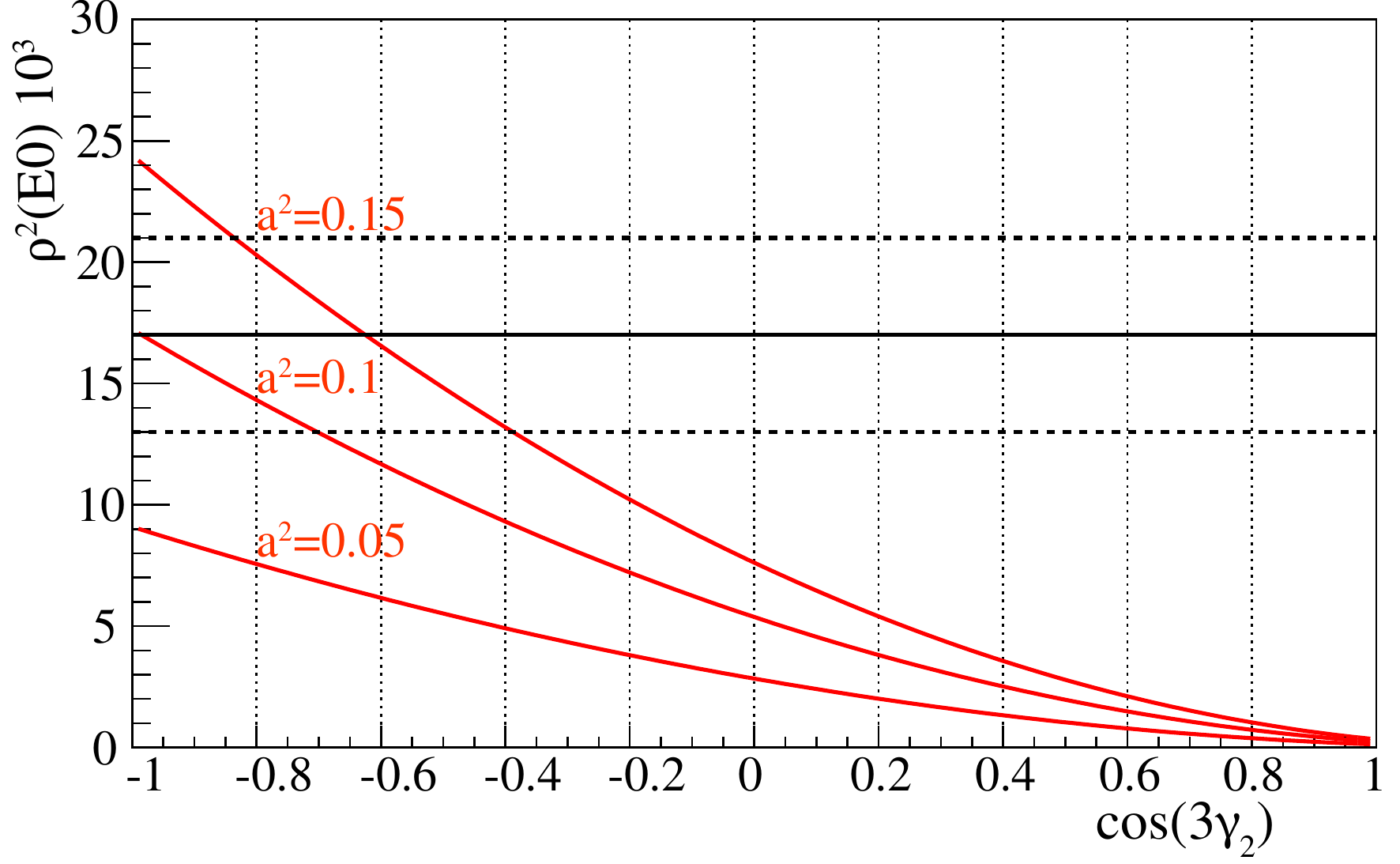}
\caption{Values of $\rho^2$ calculated as a function of the deformation
parameter $cos(3 \gamma_2)$ for different values of the squared mixing amplitude $a^2$, assumed $\gamma _1$, $\beta_1$ and
$\beta_2$ reported in Ref. \cite{svensonn}. The horizontal lines indicate the experimental value together with the $\pm \sigma$ statistical uncertainty.}
\label{gamma2}
\end{figure}
As a first step, we have considered only the terms up to the second order in $\beta$. The values of the deformation parameters $\beta^2(0_1)=0.050(2)$ and $\beta^2(0_2)=0.069(3)$ state have been extracted from the data of a Coulomb Excitation experiment performed some years ago in Ref. \cite{svensonn}. They can be expressed as a function of the unmixed $\beta_1$  and $\beta_2$ one as:
\begin{equation} \begin{split}
\beta^2(0_1) = a^2 \beta_1^2 + b^2 \beta_2^2\\
\beta^2(0_2) = b^2 \beta_1^2 - a^2 \beta_2^2
\label{beta_mixed}\end{split}
\end{equation} 

Inserting the experimental values in the Eqs. (\ref{rozeroth},\ref{beta_mixed})
the mixing coefficient $a^2$ has been calculated to be $\approx 0.1$. Since this value corresponds to a small mixing between the ground state and the $0^+_2$ state, the assumption was made that the deformations of the mixed $0_1^+$ ($0_2^+$) and unmixed $\vert1\rangle$ ($\vert2\rangle$) states are similar.


 

Under this hypothesis the value of $\rho^2(0_2^+ \longrightarrow 0_1^+)$ has been calculated keeping all the terms in Eq. (\ref{rozeroth}). The value of deformation parameter of the ground state $\gamma=20(2)^\circ$  has been taken from  Ref. \cite{svensonn}. We assume in the calculations the values for $\sqrt{\beta^2(0_1)}=0.22$, $\gamma_1=20^\circ$  and $\sqrt{\beta^2(0_2)}=0.26$ while the deformation parameter $\gamma_2$ has been varied in a reasonable range for three different set of values for $a^2$, corresponding to small mixing. The calculated values of $ \rho^2(0_2^+ \longrightarrow 0_1^+)$ are compared to the experimental one in Fig. \ref{gamma2}.
The comparison with the experimental value implies $\gamma_2 \approx 50^\circ$ for the $\vert2\rangle$ state, and hence for the $0_2^+$ state. This result would imply the coexistence of different shapes, triaxial for the ground state and oblate for the first excited $0^+$ state, in agreement with the conclusions drawn in Ref. \cite{peters}.

\section{CONCLUSIONS}
In summary, the E0 transitions in $^{106}$Pd between both $0^+$ and $2^+$ states were investigated by internal conversion electron measurements at the INFN Legnaro National Laboratories. The experiment used the newly installed SLICES setup, together with an HPGe detector. A set of K-internal conversion coefficients and monopole transition strengths was extracted.  The obtained data allow us to discriminate between the two discrepant values reported in the literature for the $\alpha_K$ of the  $2_3^+ \longrightarrow 2_1^+$ transitions. The first observation of the E0 transitions from the fourth 0$^+$ was provided but only limits on $\rho^2$(E0) values were extracted due to the limit on its lifetime.  In  $^{104}$Pd isotope hints of the existence of the fourth 0$^+$ state at 2101~keV were found re-analyzing the data of an experiment previously performed.

Calculations of the $\rho^2$(E0) values in $^{104,106}$Pd were performed in the framework of the interacting boson model, using the parameters reported in Ref. \cite{giannatiempo1998} and the monopole boson charges extracted in the present work. The agreement between theoretical results and measurements is good, once the experimental 0$^+_3$ is considered as intruder state. For both $^{104,106}$Pd isotopes predicted states having a structure resembling that of states belonging to the $n_d$=2,3 multiplets of the U(5) limit have been associated to the experimental states.
Further experimental studies aiming to give additional information about excited 0$^+$ and 2$^+$ states in the neighboring palladium isotopes are necessary to establish their interpretation as lying within the IBA-2 model space.

The experimental value of the $\rho^2(E0; 0_2^+ \longrightarrow 0_1^+)$  has been also compared to that calculated in a simple two-states mixing model, to obtain further insights on the mixing and deformation of the first two $0^+$ states in this nucleus.

\section{ACKNOWLEDGEMENTS}
The authors would like to thank the staff of the CN accelerator (LNL) for providing the beams used in this experiment, M.~Loriggiola for producing the targets, and the mechanical workshops of the INFN divisions of Florence and the University of Camerino for their contribution.
E. R. G. wishes to acknowledge
the Centro E. Fermi for financially supporting his postdoctoral fellowship
through the project BESTRUCTURE.

\bibliography{biblio}

\end{document}